\documentclass[pre,preprint]{revtex4-1} 

\usepackage{amsmath}  
\usepackage{amsfonts} 
\usepackage{graphicx} 
\usepackage{subfig}

\begin{document}


\title{Inducing Self-Organized Criticality in a network toy model by neighborhood assortativity}

\author{Alfonso Allen-Perkins}
\email{alfonso.allen@hotmail.com} 
\affiliation{Complex System Group, Technical University of Madrid, 28040 Spain}

\author{Javier Galeano}
\email{javier.galeano@upm.es}
\affiliation{Complex System Group, Technical University of Madrid}

\author{Juan Manuel Pastor}
\email{juanmanuel.pastor@upm.es}
\affiliation{Complex System Group, Technical University of Madrid}


\date{\today}

\begin{abstract}
Complex networks are a recent type of frameworks used to study complex systems with many interacting elements, such as Self-Organized Criticality (SOC). The  network nodes's tendency to link to other nodes of similar type is characterized by assortative mixing. Real networks exhibit assortative mixing by vertex degree, however typical random network models, such as Erd\"{o}s-R\'{e}nyi or Barab\'{a}si-Albert, show no assortative arrangements. In this paper we introduce the \emph{neighborhood assortativity} notion, as the tendency of a node to belong to a community (its neighborhood) showing an average property similar to its own.
Imposing neighborhood assortative mixing by degree in a network toy model, SOC dynamics can be found. These dynamics are driven only by the network topology. The long-range correlations resulting from the criticality have been characterized by means of fluctuation analysis and show an anticorrelation in the node's \emph{activity}.
 The model contains only one parameter and its statistics plots for different values of the parameter can be collapsed into a single curve. The simplicity of the model allows performing numerical simulations and also to study analytically the statistics for a specific value of the parameter, making use of the Markov chains. 

\end{abstract}

\maketitle 

\section{Introduction} 

Self-Organized Criticality is a paradigm of complex system.
In their seminal work, Bak, Tang and Wiesenfeld  (1987) introduced the idea of self-organized criticality (SOC) using a computer cellular automaton as a sandpile experiment \cite{Bak87}.  Their system assembled itself in a critical state. When the system relaxed (recovering the stationary state) it showed spatial and temporal self-similarities. 
Systems exhibiting SOC dynamics are open dissipative systems,  involving two time scales: a slow energy income and a quick relaxation. Empirical examples that have been linked to SOC dynamics are earthquakes \cite{Pisarenko03}, solar flares \cite{Lu91}, neuronal activity \cite{Beggs03}, or sand piles \cite{Held90, Grumbacher93}  among others. 

In order to determine the  physical properties of these dynamics, different models have been proposed \cite{Pruessner12}. The archetypal model of SOC  is the \textit{sandpile} model  which mimics the process of adding sand grains one by one  over a sand pile \cite{Bak88}. The mechanical instability is simulated by a threshold height (or  height difference relative to its neighbors). This process allows to develop \textit{avalanches} with event size distribution similar to those of the sand pile experiments. In order to model earthquakes SOC dynamics Olami \textit{et al.} (1992)   introduced a non-conservative SOC model (OFC model) based on 2D spring-block system connected to a rigid driving plate \cite{Olami92}. Their cellular-automaton
displayed similar statistics and gave a good prediction of the Gutenberg-Richter law. Caruso \textit{et al.} (2007) analyzed the OFC model in regular lattice and small world network \cite{Caruso07}. They reported a well-defined power-law distribution of the avalanche size and characterized the presence of criticality by the PDF of the differences between avalanche sizes at different times ($t$ and $t+\Delta t$).
 
The study of complex systems employing Network Science framework has attracted much interest in many interacting-elements systems \cite{Caldarelli09}. 
Several models for studying SOC on complex networks have been proposed \cite{Arcangelis01, Moreno02, Bianconi04}.
In these models criticality is produced by a ``fitness'' parameter defined on the nodes or by a rewiring process \cite{Bianconi04, Fronczak06}.

We present a simple network model that mimics the instability condition of the sandpile models imposing a local stability condition associated to an average property of its neighborhood. This neighborhood assortativity produces SOC dynamics driven by the network's topology, namely, a node will become unstable when its degree is greater than a threshold condition (like in sandpile models). This generalization of sandpile models can describe many interacting-elements system in which the maximum value of the node's property depends on its neighborhood's values. As far as we know this is the only SOC model that is driven exclusively by the network topology (neither rewiring requirements nor non-topologycal properties associated to the nodes have been used).

The model definition and its main features are explained in Section \ref{Sec:Model}. Numerical simulations are reported in Section \ref{Sec:Results}. In the first part of Section \ref{Sec:Results}, we characterize the neighborhood topology conducted by the stability condition (introducing the \emph{neighborhood assortativity}). In the second part, we show a complete characterizations of SOC dynamics and we compare our results with the classical OFC model \cite{Caruso07}.
In Section \ref{Sec:statistics} we develop the algorithms for the probability distribution by means of the Markov chains for a special case where the network is restricted to a linear chain, and compare the results with numerical simulations. Conclusions are summarized in Section \ref{Sec:Conclusion}.

\section{Toy Model}
\label{Sec:Model}

Starting from a single node the network grows by adding a new node (with a single link) at each time step,
nonetheless a topological stability condition constrains the growth.
After  a new addition the system can result in an unstable configuration. This unstable state leads to a relaxation process which, eventually, can end with the removal of nodes.
The interplay between dynamics and topology drives the system.

In this model the stability of a node depends on the ``support'' of its neighbors as follows: the node's degree must be less than or equal to the average degree of its neighbors plus a global constant (hereafter \textit{buffering capacity}, $C$). Therefore the stability condition can be written as follows:
\begin{equation}
k_{i} \leq \left\langle k_{j} \right\rangle_{N_{i}} + C 
\label{eq:stabil_cond}
\end{equation}
\noindent where $k_{i}$ is the $i$-node's degree, and $N_{i}$ is the set of nearest neighbors of node $i$. 

This inequation may be rewritten as an equation introducing a new local parameter $ a_{i} $:
\begin{equation}
k_{i} + a_{i} = \frac{1}{k_{i}}\sum_{j\in N_{i}}k_{j} + C
\label{eq:stabil_cond_alpha}
\end{equation}
\noindent where $a_{i}$ is a grade of stability, i.e., the higher $a_{i}$ the more stable the node is. We will identify the term $-a_{i}$ as the node's \emph{activity}. Note that $a_{i}$ is negative for unstable nodes.

When a node becomes unstable, one of its $k_{i}$ links is randomly removed and the smallest subnet is deleted. Since the degree has changed, the stability conditions of the node and its neighbors have to be checked again in an iterative process until every node in the network is stable.
The set of removals nodes performed until every node in the network is stable represents an \textit{avalanche}. The \textit{size} of the avalanche can be defined as the total number of nodes removed from the network.

Starting from a single node or a small network these dynamics make the system evolve towards a finite network whose average size, in the stationary regime, depends on the \textit{buffering capacity} constant.
Note that this model generates \textit{tree-like} networks since one link is added at each time step and there is no rewiring between nodes, thus there are no cycles in the network. Moreover, starting from a full-connected network as initial seed after a long time period every cycle will break up (for any finite $C$). Therefore, the average number of links is always less than $2$. These tree-type networks can be found in the tree topology of physical connections on a LAN (hybrid  bus and star topologies), where a switch can only distribute through a limited number of connections, and also in branching processes such as fractal trees \cite{Vandewalle97}.

The node's stability condition is related to the average degree of its neighbors. Thus, it will produce networks of positive \textit{neighborhood assortativity}, i.e., the node's degree tends to be similar to the average degree of its neighborhood, as it is shown in the next Section.

This condition implies a maximum degree in the network. This maximum degree depends on the constant $C$. 
Starting from a single seed, the maximum degree of the network can be increased when the  new node is added to a node with the highest degree but connected to neighbors with the same maximum degree. In this case the new stability condition can be written as:
\begin{equation}
k_{i}+1 \leq \frac{k_{i}^ 2 +1}{k_{i}+1} + C 
\label{eq:stabil_cond_max}
\end{equation}

The minimum value of $C$ satisfying $k_{i}+1=k_{max}$ will be:
\begin{equation}
C_{min}^{k_{max}}=\frac{2\left(k_{max}-1\right)}{k_{max}}
\label{eq:kmax}
\end{equation} 

For example, for $k_{max}=2$, $C_{min}=1$; for $k_{max}=3$, $C_{min}=4/3$; and for $k_{max}=\infty$, $C_{min}=2$.

Dynamically, this model shows four behaviors depending on the global parameter $C$: i) $0\leq C <1$, whose solution is the trivial \textit{duple} since adding a new node is always unstable; ii) $1\leq C <4/3$, generates only  linear chains, i.e., the possible stable configurations require $k_{max}=2$ (this fact will allow us to study statistically this case  in Section \ref{Sec:statistics});  iii) $ 4/3 \leq C < 2 $, produces networks with the value of $k_{max}$ restricted by Eq.~\ref{eq:kmax}; however, the average size of these networks is limited to a value dependent on $C$ (as it can be seen in Section \ref{Sec:Results}, Figure \ref{fig2:time_evol_N}); iv) $C \geq 2$ produces networks without any limit in the maximum degree (but also with the stationary average size limited to a value dependent on $C$). The limit case $C=\infty$ allows adding a new node to any node in the network; all configurations are stable, thus there won't be any pruning events and the network will grow without any limit.

\section{Numerical Results}
\label{Sec:Results}

\begin{figure}[h!]
\centering
\includegraphics[width=120mm]{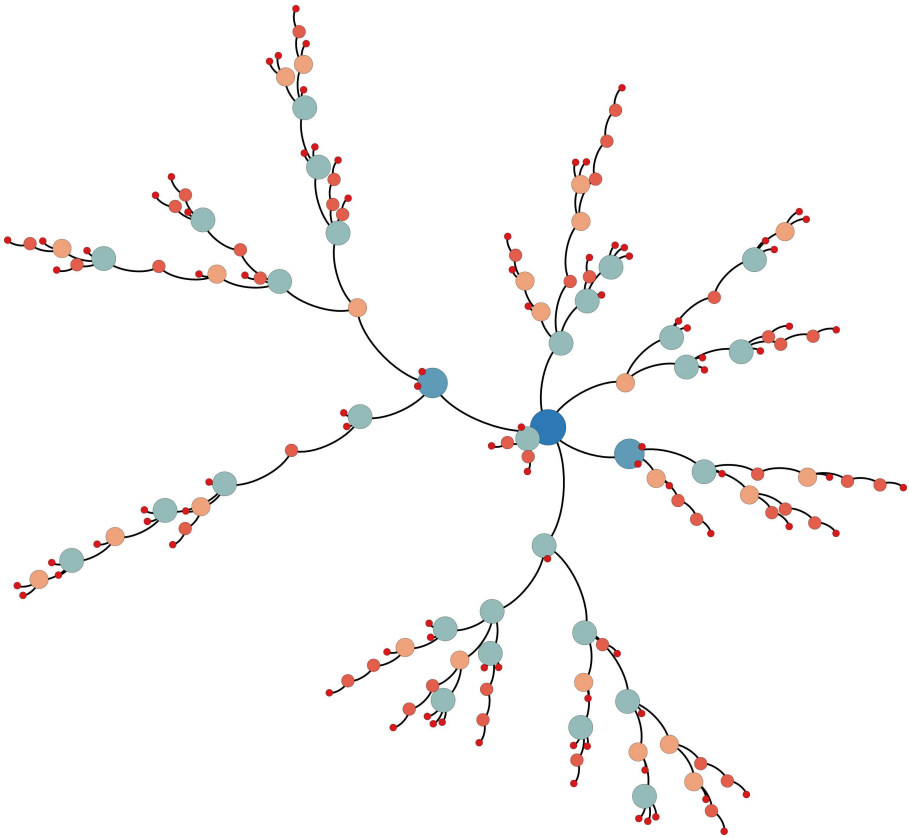}
\caption{Network snapshot for $C=2.5$, at time $t=3000$. Node's degree is size-coded, from $k=1$ to $k=6$.}
\label{fig1:network_C2p5}
\end{figure}

We performed numerical simulations starting from a single node and checked that by starting from a different number of nodes, only the statistics at initial time steps change while at the stationary state they stay the same. A snapshot of a network with $C=2.5$ at time $t=3000$ is depicted in Fig.~\ref{fig1:network_C2p5}.

The stability condition in the model implies a new type of assortative mixing in which the nodes's tendency to link does not depend on its nearest neighbors' property but on the neighborhood's average property.
In order to characterize the \emph{neighborhood's assortativity} we have assigned 
a new property to the node: the neighborhood's average degree, 
defined as, $\tilde{k}_{i}\equiv \sum k_{j}/(k_{i}+1)$, where $j\in N_{i}$, i.e, the average degree including the $i-$node.

For this magnitude we can obtain an assortativity coefficient as the standard Pearson correlation coefficient. The average value of the neighborhood assortativity coefficient, averaged over $1000$ networks at the stationary state, is around $r_{\tilde{k}} = 0.45$, for $C$ in the range $[2.0 - 3.0]$. The significance of these values is evaluated by the \textit{jackknife method} \cite{Newman03}, with an estimated error of $\sigma_{r} = 0.05$.  This new approach of assortative mixing by a neighborhood's average property indicates that a node is more likely to connect to a \emph{neighborhood} whose average degree is close to its own degree.

It is worth noting that assortative mixing by vertex degree is null ($r=0$) 
for this model as for random graphs (Erd\"{o}s-R\'{e}nyi, E-R) or the Barab\'{a}si-Albert (B-A) model \cite{Newman02}. However, assortative mixing by \textit{neighborhood's average degree} is significantly positive, $r_{\tilde{k}}=0.45$, while it is null for E-R or B-A models. 

Our preliminary results on neighborhood assortativity for real-world and synthetic networks (like trimmed and diluted Cayley trees \cite{Vandewalle97})  show that some of them exhibit positive neighborhood asortativity and null degree assortativity.

Regarding the dynamics, this model includes two time scales, a ``slow'' energy income (one node per time unit, $t$) and a ``fast'' relaxation process triggered by the stability criterion (associated to a micro-time). The instability of a node involves the pruning of a link, which can lead to another instability
situation at the next ``micro-time'', and so on, producing eventually an avalanche. These fluctuations
of slow increase and sudden decrease can be seen as a sign of criticality; the system is in a marginal stability state and evolves itself toward the stationary state. A complete characterization of its SOC dynamics is presented below.

\begin{figure}[h!]
\centering
\includegraphics[width=\textwidth]{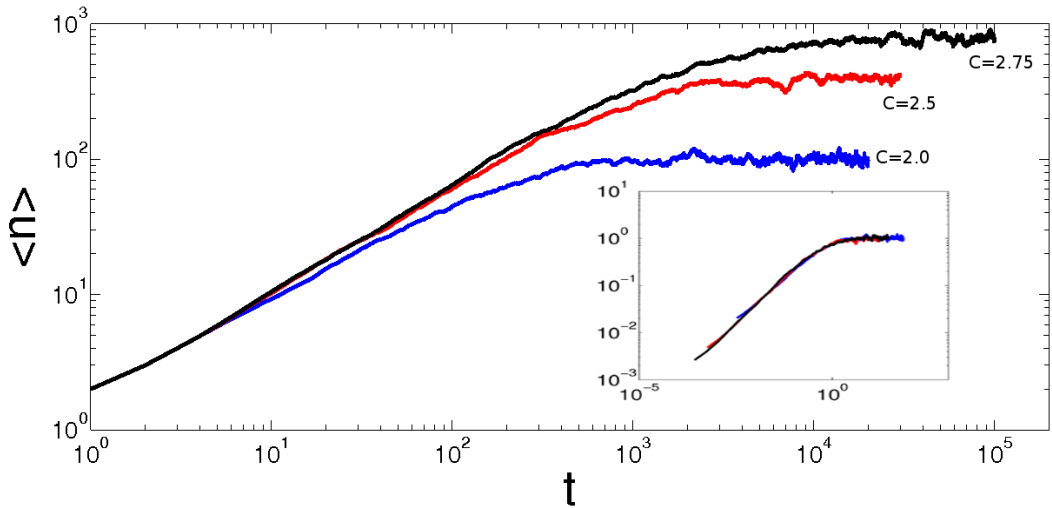}
\caption{ Average time evolution of the total number of nodes for different \textit{buffering capacity} constants (color online): $C=2.0$ (blue line), $C=2.5$ (red line), and $C=2.75$ (black line). Inset: Collapse of the previous plots.}
\label{fig2:time_evol_N}
\end{figure}

Time evolution of the system size (number of nodes) averaged over $100$ realizations, for different values of the \textit{buffering capacity}  constant is plotted in Fig.~\ref{fig2:time_evol_N}. The marginal stability of SOC state can be determined by the stationarity of the averaged value of a characteristic global magnitude, like the system size (or the average degree), as can be observed in Fig.~\ref{fig2:time_evol_N}.
These average sizes at stationary state ($n_{stat}$) versus \textit{buffering capacity} constant can be fitted by a power-law (when $k_{max}$ is unlimited). Thus, statistics from different values of $C$ can be collapsed after rescaling by $n_{stat}$. The transition time can also be fitted with respect to the average stationary system size as $t_x \sim n_{stat}^{1/z}$, with an exponent $z=0.8 \pm0.1$ (following the notation of fractal growth dynamics). 
In the inset of Fig.~\ref{fig2:time_evol_N} we represent the master curve for the main figure (collapsed from the different plots).

\begin{figure}[h!]
\centering
\subfloat[]{\label{fig3a:time_evol_N_C2}\includegraphics[width=0.5\textwidth]{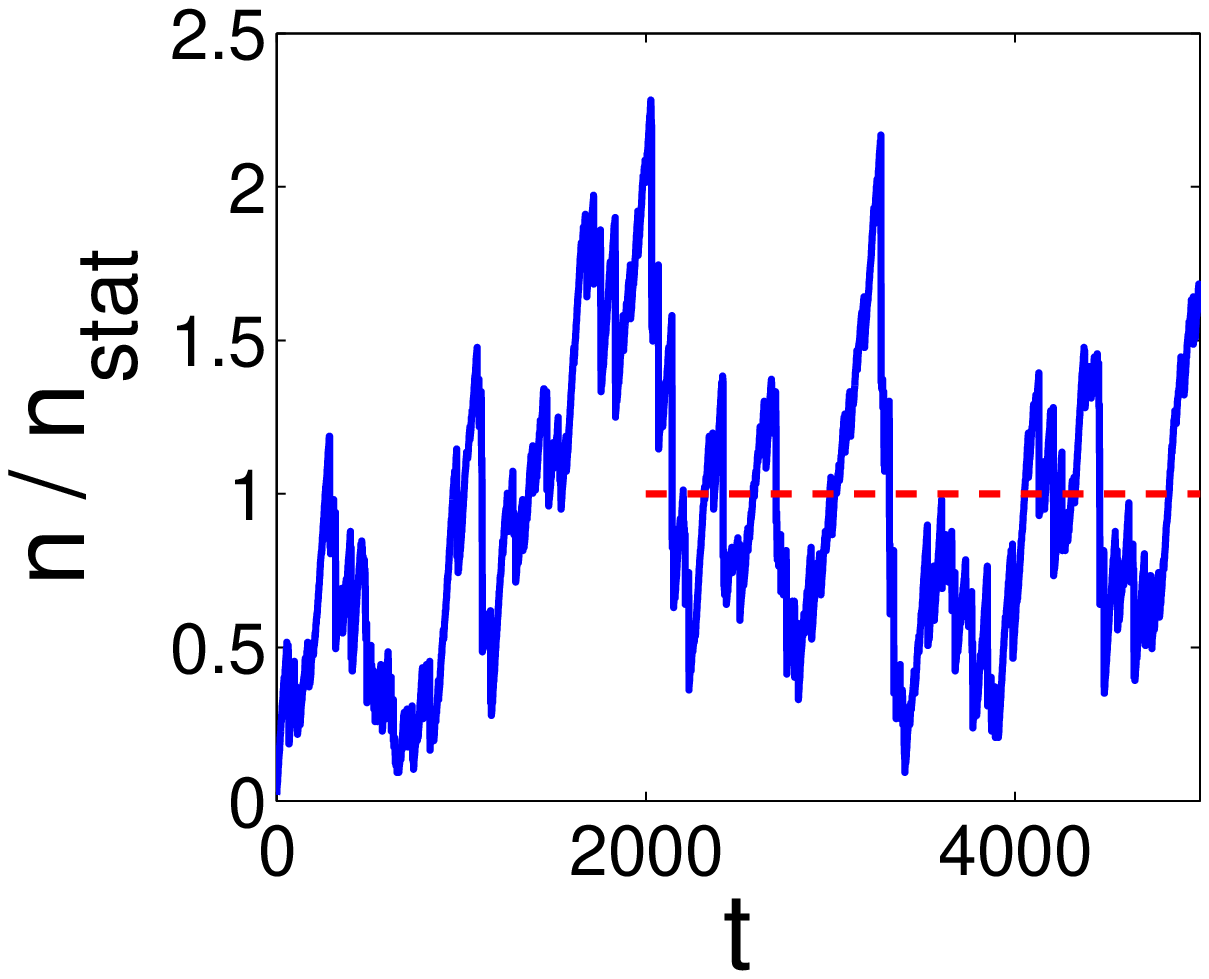}}
\subfloat[]{\label{fig3b:power_spectr}\includegraphics[width=0.5\textwidth]{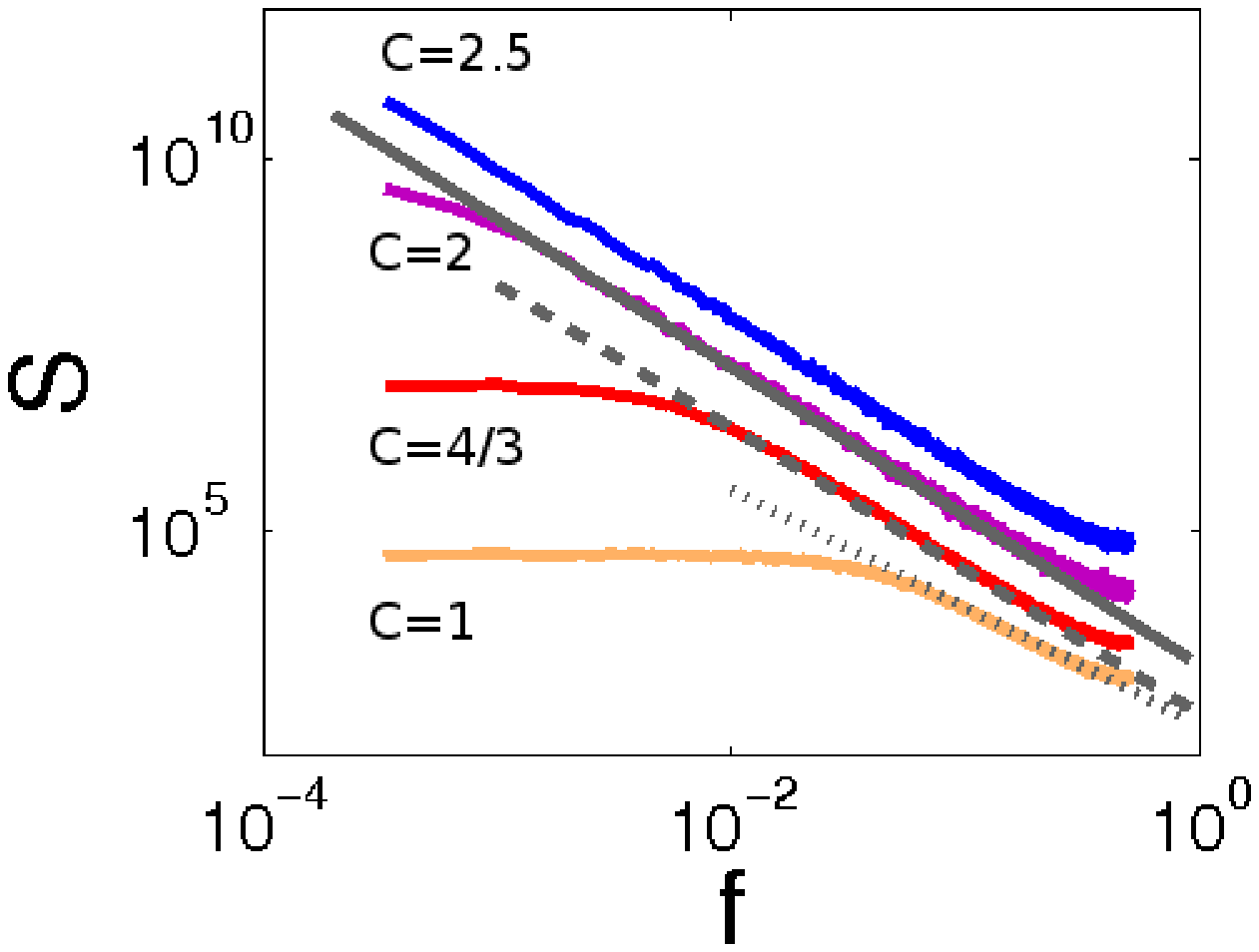}}
\caption{a) Time evolution of the network size normalized by the stationary system size, for $C=2$; horizontal dashed line represents the average stationary system size. b)Power spectra (color online), for $C=1$ (orange), $C=4/3$ (red), $C=2$ (purple), $C=2.5$ (blue); gray straight lines of slope $m$ are guides to the eye: $m=-1.6$ dotted line, $m=-1.9$ dashed line, and $m=-2.0$ continuous line.}
\end{figure}

A typical time evolution of the normalized number of nodes, $n(t)/n_{stat}$, (for $C=2$) is plotted in Fig.~\ref{fig3a:time_evol_N_C2}, where the dotted horizontal line corresponds to the average stationary system size. 
Fig.~\ref{fig3b:power_spectr} shows the power spectra of these discharge events for different values of $C$ (averaged over 1000 realizations). For $C\geq 2$ (dynamics with no restrictions on $k_{max}$) a well-defined $f^{-2}$ fit is obtained, in agreement with mass fluctuations in sandpile experiments \cite{Held90}, and theoretical results \cite{Jensen89}. When $1\leq C<2$ the power spectra exhibits a lower slope and a cutoff at low frequencies.

\begin{figure}[h!]
\centering
\includegraphics[width=0.8\textwidth]{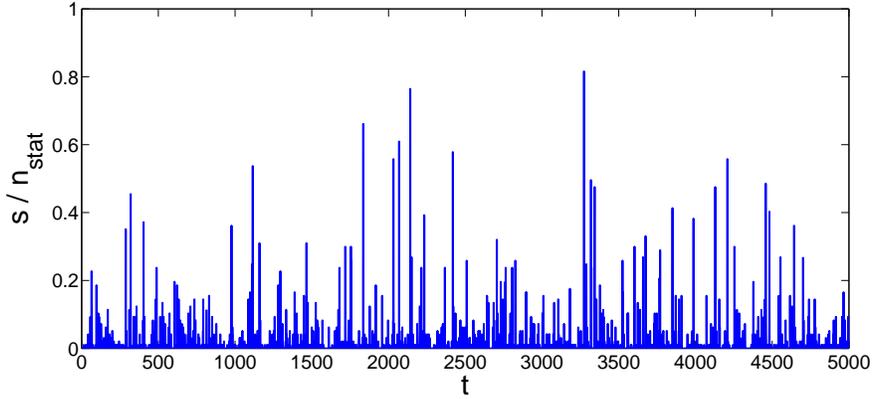}
\caption{Normalized avalanche sizes evolution during $5000$ time units for $C=2$.}
\label{fig4:time_evol_event_C2}
\end{figure}

\begin{figure}[h!]
\centering
\subfloat[]{\label{fig5a:distr_size_event}\includegraphics[width=0.49\textwidth]{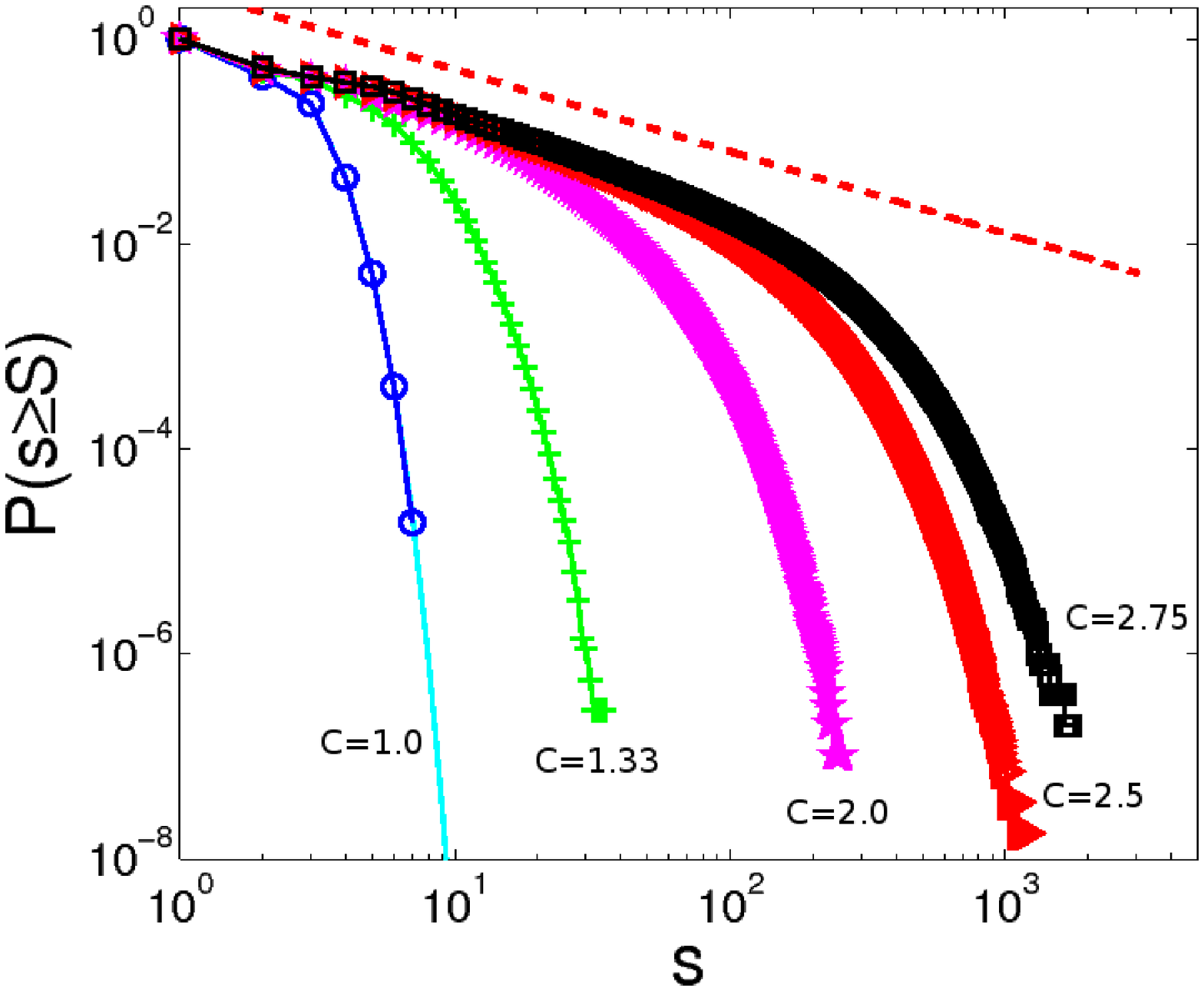}}
\subfloat[]{\label{fig5b:colap_distr_size_event}\includegraphics[width=0.51\textwidth]{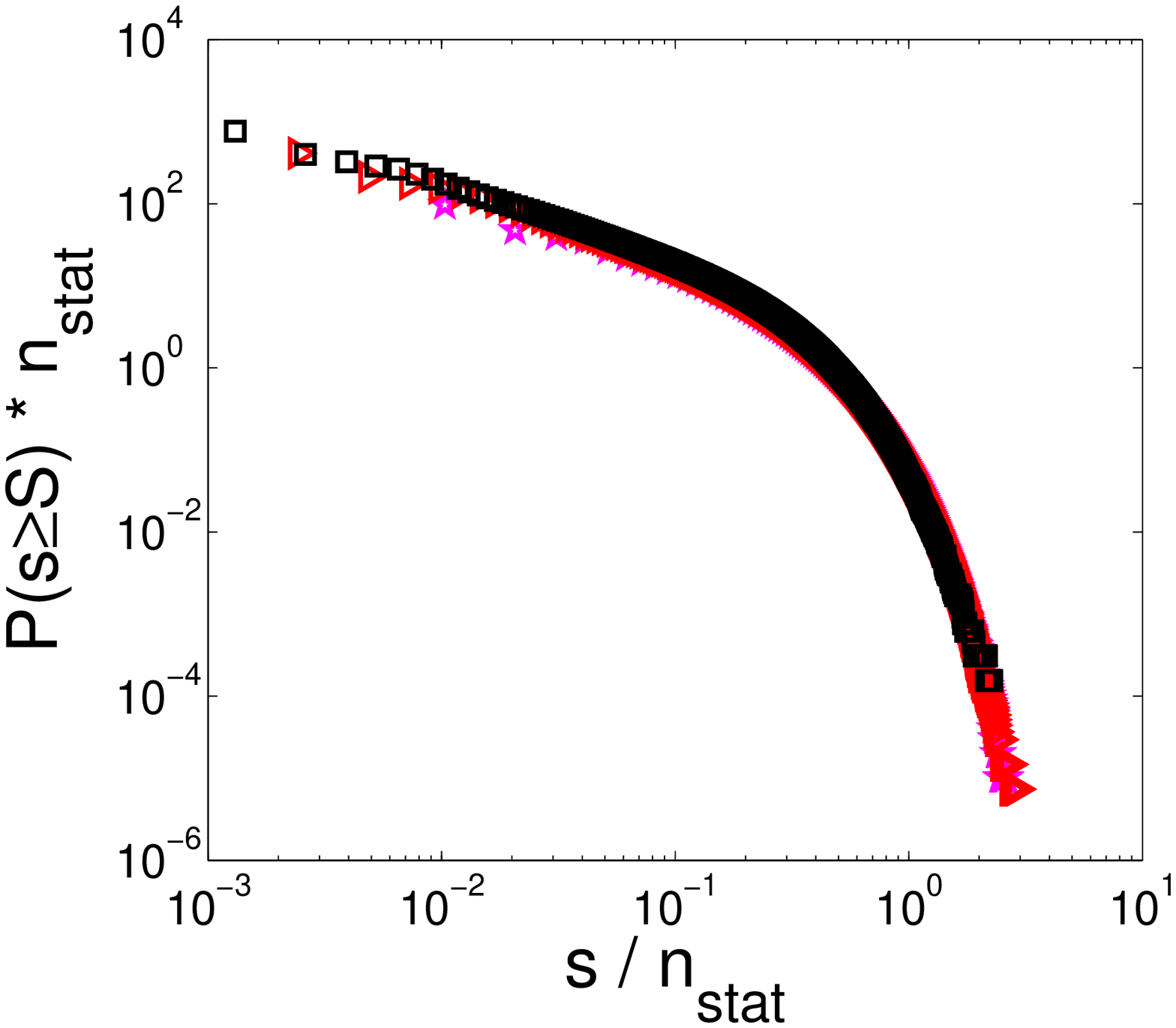}}
\caption{ a) Cumulative normalized event size distribution for different \textit{buffering capacity} constants (color online):  $C=1$ (theoretical) cyan line, $C=1$ blue circles, $C=4/3$ green pluses, $C=2.0$ violet stars, $C=2.5$ red triangles, and $C=2.75$ black left triangles; the dashed line with slope $\gamma = -0.8$ is a guide to the eye for the power-law regime. b) Collapse of the same plot for $C\geq 2$.}
\end{figure}

A characteristic aspect of SOC is the cumulative avalanche size distribution that exhibits power-law scaling with finite-size effect.
Fig.~\ref{fig4:time_evol_event_C2} depicts the normalized avalanche sizes evolution during 5000 time units, for $C=2$, where Y-axis represents the size of events $s$ normalized with respect to the stationary  system size, i.e., $s^*=s/n_{stat}$ (Note that in these units avalanches of size greater than 0.5 can be observed).
The cumulative avalanche size distribution for different values of $C$  exhibits a power-law behavior $P(s\geq S)\sim s^{-\gamma}$.
The scale-free nature of the event size distribution allows us to collapse all the event size distributions for different \textit{buffering capacity} values, by normalizing with the stationary system size. 
The normalized avalanche size cumulative distribution for different \textit{buffering capacities}, from $C=1$ to $C=2.75$ is plotted in Fig.~\ref{fig5a:distr_size_event}; the larger the \textit{buffering capacity}, the wider the range of the power-law behaviour. In Section \ref{Sec:statistics} we obtain theoretically the same results for $C=1$ (cyan solid line in Fig.~\ref{fig5a:distr_size_event}). The collapse of plots for $C\geq 2$ is depicted in Fig.~\ref{fig5b:colap_distr_size_event}. The cumulated power-law exponent $\gamma = -0.8 \pm 0.1$ is in agreement with the corresponding value for the OFC model ($-1.8$ exponent for the PDF)\cite{Olami92}.

\begin{figure}[h!]
\centering
\subfloat[]{\label{fig6a:WTD}\includegraphics[width=0.5\textwidth]{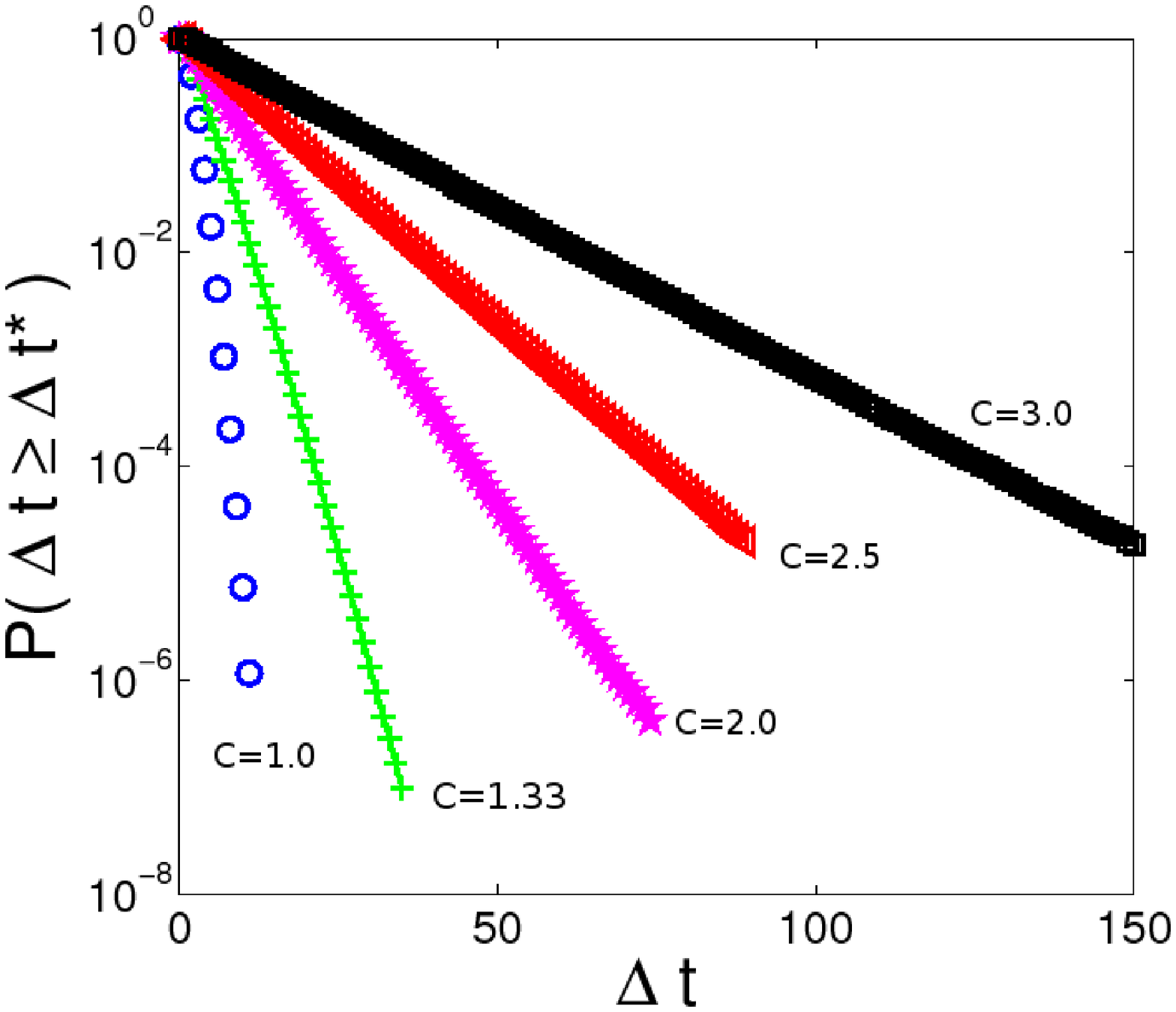}}
\subfloat[]{\label{fig6b:collap_WTD}\includegraphics[width=0.51\textwidth]{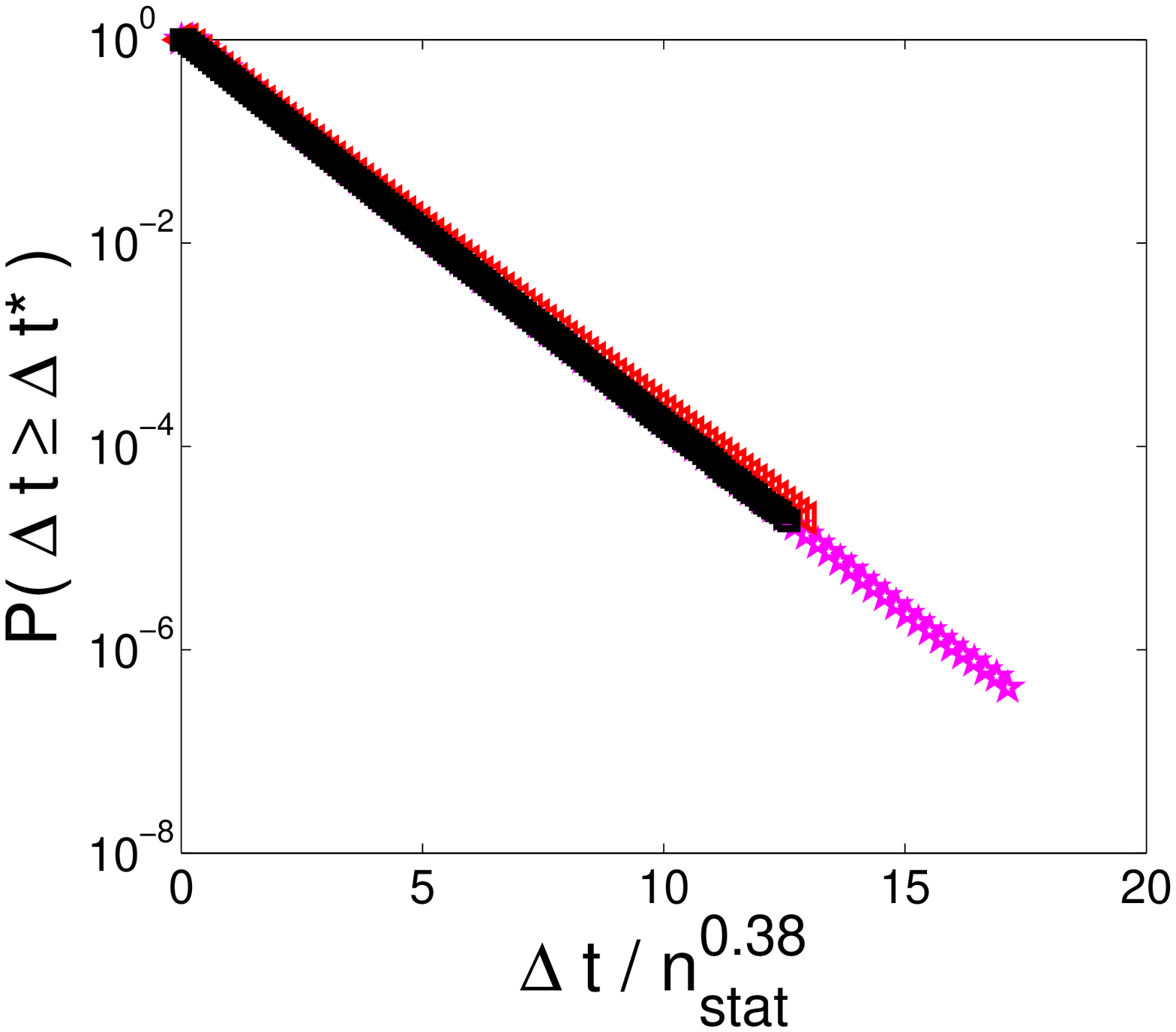}}
\caption{a) Cumulative WTD for different \textit{buffering capacity} constants (color online): $C=1.0$ (blue circles), $C=1.33$ (green pluses), $C=2.0$ (pink stars), $C=2.5$ (red triangles), and $C=3.0$ (black squares), in log-lin scales.  b) Collapse of plots on the left according to Eq.~\ref{eq:t_c}, for $C \geq 2$, with $\beta=0.38$.}
\end{figure}

Another magnitude studied in SOC time series is the \textit{waiting time distribution} (WTD) which represents the distribution of time intervals between two consecutive events. In generic SOC models, when the triggering is not correlated \cite{Boffetta99,Sanchez02}, the distribution shows an exponential behavior, unlike earthquakes or solar flares activity data with a power-law distribution. Fig.~\ref{fig6a:WTD} shows the cumulative WTD for different \textit{buffering capacities} from $C=1.0$ to $C=3.0$. Note the log-scale in the Y-axis and linear-scale in the X-axis. The cumulative WTD can be fitted by an exponential function $exp(-\Delta t / t_{C}^{\ast})$. 
We can collapse the curves for $C\geq 2$, with different exponential slopes (Fig.~\ref{fig6b:collap_WTD}), rescaling the interval times with a power of the stationary system size,
\begin{equation}
t_{C}^{\ast} \sim n_{stat}^{\beta}
\label{eq:t_c}
\end{equation}

\noindent with an exponent $\beta=0.38 \pm 0.05$.

Referring to the correlations statistics, different procedures have been reported in the literature \cite{Markovic14, McAteer2015}. Based on the idea of the relative difference in the size of avalanches described in studies of solar flares or earthquakes, we can define the relative difference in the size of avalanches at different times, $t$ and $t+\Delta t$, as follows:
\begin{equation}
\delta S_{\Delta t} = \frac{s(t+\Delta t) - s(t)}{\sigma_{\Delta t}}, \qquad \sigma_{\Delta t}^{2}=\left\langle\left( s(t+\Delta t) - s(t) \right)^{2}\right\rangle 
\end{equation}

\begin{figure}[h!]
\centering
\includegraphics[width=0.8\textwidth]{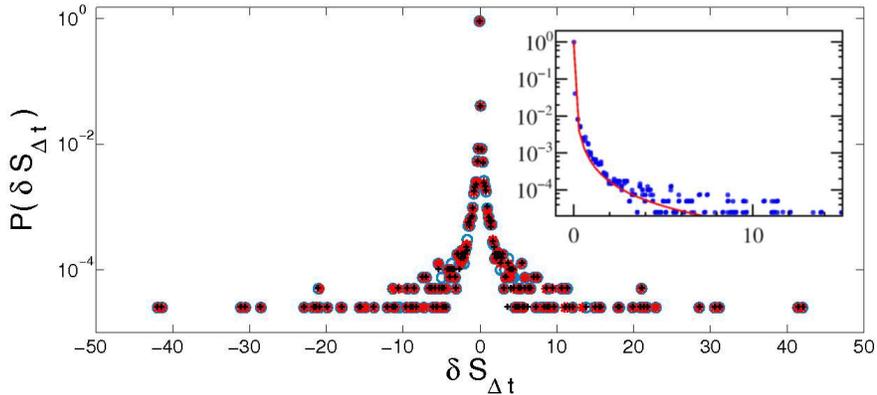}
\caption{Probability distribution of removed nodes fluctuations, for different time intervals $\Delta t=1$ (blue circles), $10$ (red asterisks),  $1000$ (black pluses). Inset: zoom of positive values and q-Gaussian fit for $\Delta t=1, 10, 100, 1000$, with $\delta s<20$, and $q=2.2$}
\label{fig7:prob_distr_energy}
\end{figure}

The probability distributions  for removed nodes  fluctuations $P(\delta S_{\Delta t})$, obtained for different values of $\Delta t$, from $\Delta t=1$ to $\Delta t=1000$ are depicted in Fig.~\ref{fig7:prob_distr_energy}. 
The overlapping of different inter-event scales $\Delta t$ indicates the lack of time scales in the correlations.

Caruso \textit{et al.}  (2007) reported that in the critical Olami-Feder-Christensen model (on a small world topology) the PDF of the avalanche size differences (referred as "returns") can be fitted by a \emph{q-Gaussian} curve 
$ f(x)=A\left[1-(1-q)x^{2}/B\right]^{1/(1-q)}$  \cite{Caruso07}. In the critical regimen they obtained a value of $ q=2.0\pm 0.1 $.
In our case, similar results were obtained.
Inset of Fig.~\ref{fig7:prob_distr_energy} shows a zoom for small ``returns'' and a suitable fitting for different values (from $ \Delta t =1$ to $\Delta t = 1000$); solid line corresponds to the fitting by a \emph{q-Gaussian} curve, with an exponent $ q=2.2 \pm 0.2$,  compatible with   the OFC model \cite{Caruso07}.

Finally, we have tried to quantify the long-range spatial correlations following the fluctuation analysis introduced by Rybski \textit{et al.}  (2010) for networks \cite{Rybski10}. This approach  is based on the fluctuations of degree sequence along shortest paths of  length $d$, and it can be adapted to any topological property of the network, like a node activity. In our case, the magnitude playing the role of node activity is  $-a_{i}$ defined in Eq.~\ref{eq:stabil_cond_alpha} (the more negative $a_{i}$ the more unstable the $i$-node is).

\begin{figure}[h!]
\centering
\includegraphics[width=0.85\textwidth]{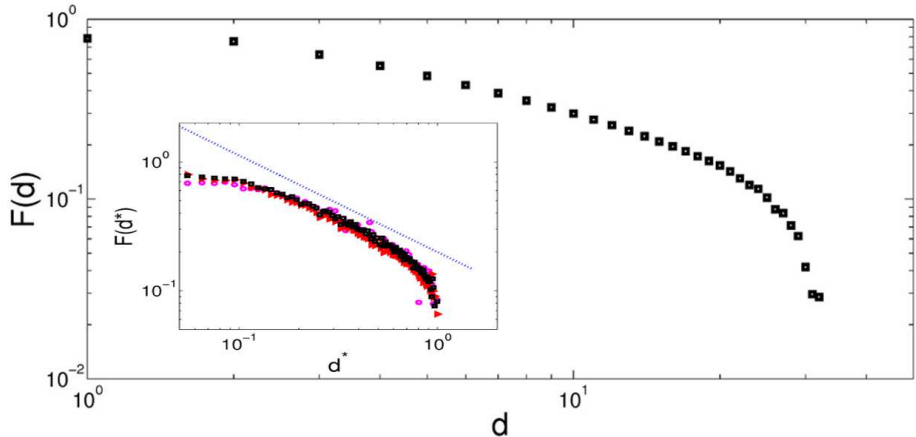}
\caption{ Fluctuation function $F(d)$ for $C=2.75$. Inset:  $F(d^{*})$ functions for $C=2$ (pink circles), $C=2.5$ (red triangles), and $C=2.75$ (black squares); the dashed line is an eye-guide with slope $-0.7$.}
\label{fig8:F_d}
\end{figure}

Following the procedure described in \cite{Rybski10}, we have considered all the shortest paths of length $d$ in the network and calculated the standard deviation of the averages  of our activity $a_{i}$, $F(d)$. 
Figure \ref{fig8:F_d} shows the fluctuation function $F(d)$ for different values of $C$, averaged over $2000$ snapshots. A power-law tendency can be observed (superimposed with the exponential finite-size effect). Since the usual Hurst-like exponent $\alpha_{H}$ is related with the fitted value $\alpha$ by $\alpha_{H}=\alpha + 1$ \cite{Rybski10}, positive long-range correlations are characterized by exponents $-1/2 < \alpha < 0$, while the negative ones are characterized by exponents $-1 < \alpha < -1/2$. 

The main challenge of studying the networks of this model is the variation of their size. In order to overcome this difficulty we have rescaled the distance $d$ with the diameter of the network, $d^{*}=d/d_{max}$. Inset of Figure \ref{fig8:F_d} depicts the fluctuation function for three values of $C$ by rescaling the distance, $F(d^{*})$.
For different values of $C$, we have always obtained an exponent $\alpha < -0.5$, indicating anticorrelations in the activity. This result is in agreement with the meaning of node's activity ($-a_{i}$) since a node with high activity (more negative values of $a_{i}$) has a higher degree than its neighbors (in average).

This anticorrelation can be also found employing the assortative mixing by activity, $r_{a}$. As mentioned by Newmann (2003) one can compute the standard Pearson correlation coefficient for any scalar variable associated to the nodes \cite{Newman03}. The value of this correlation coefficient for the activity is $r_a = - 0.5 \pm 0.1$ (where the error is calculated by the jackknife method \cite{Newman03}) in agreement with the result shown by the fluctuation function. It is worth remembering that assortative mixing by vertex degree is null like in the case of other typical random network models.

\section{Statistical results}
\label{Sec:statistics}
 In this model, there exists a special case that, due to its simplicity, can be treated statistically.
When the \textit{buffering capacity} constant is $1\leq C < 4/3$ the node degree is limited to $k_{max}=2$. Therefore, the only possible result is a linear chain whose size evolves stochastically.

\begin{figure}[h!]
\centering
\subfloat[]{\label{fig9a:evol_prop_size_C1}\includegraphics[width=0.5\textwidth]{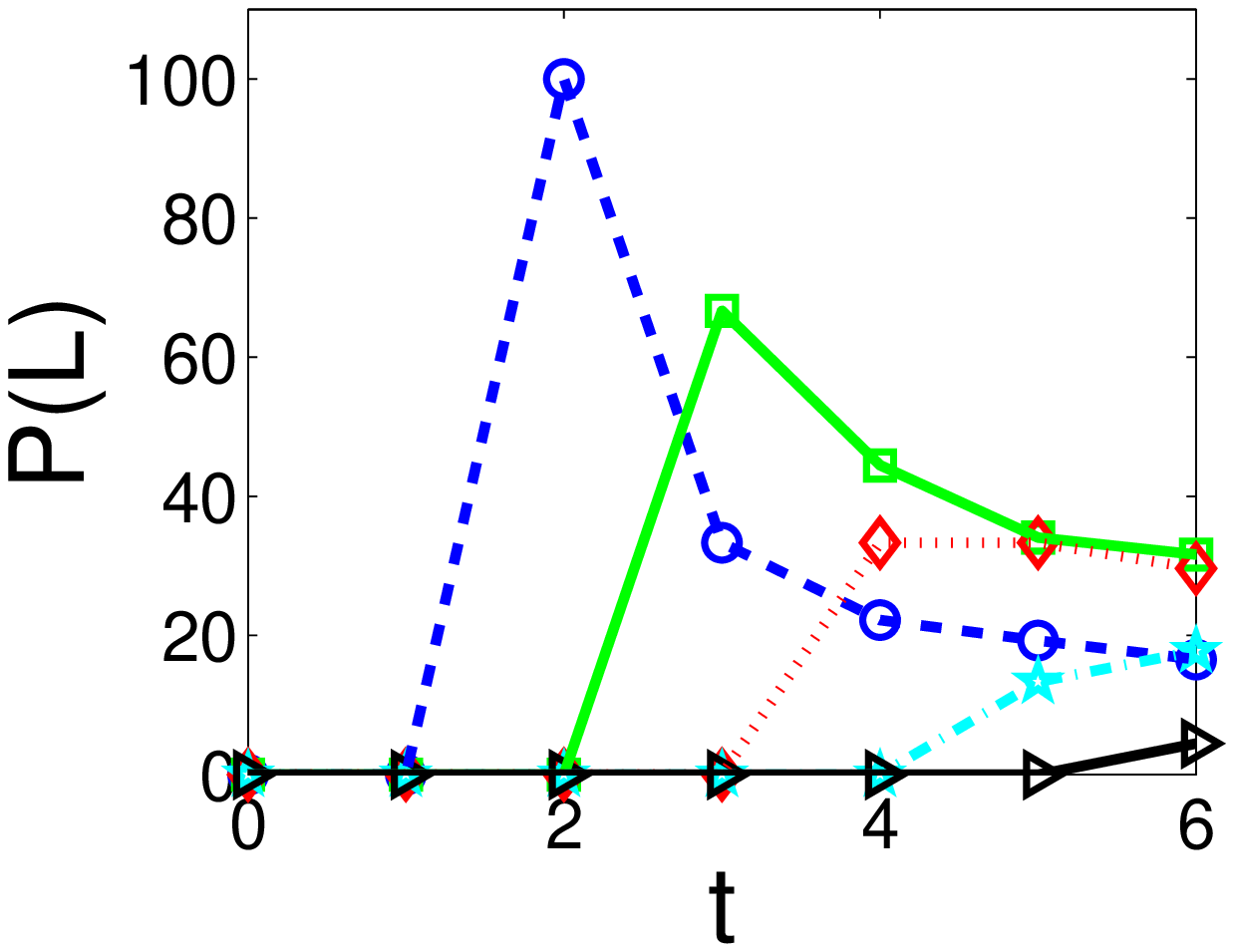}}
\subfloat[]{\label{fig9b:evol_event_distr}\includegraphics[width=0.5\textwidth]{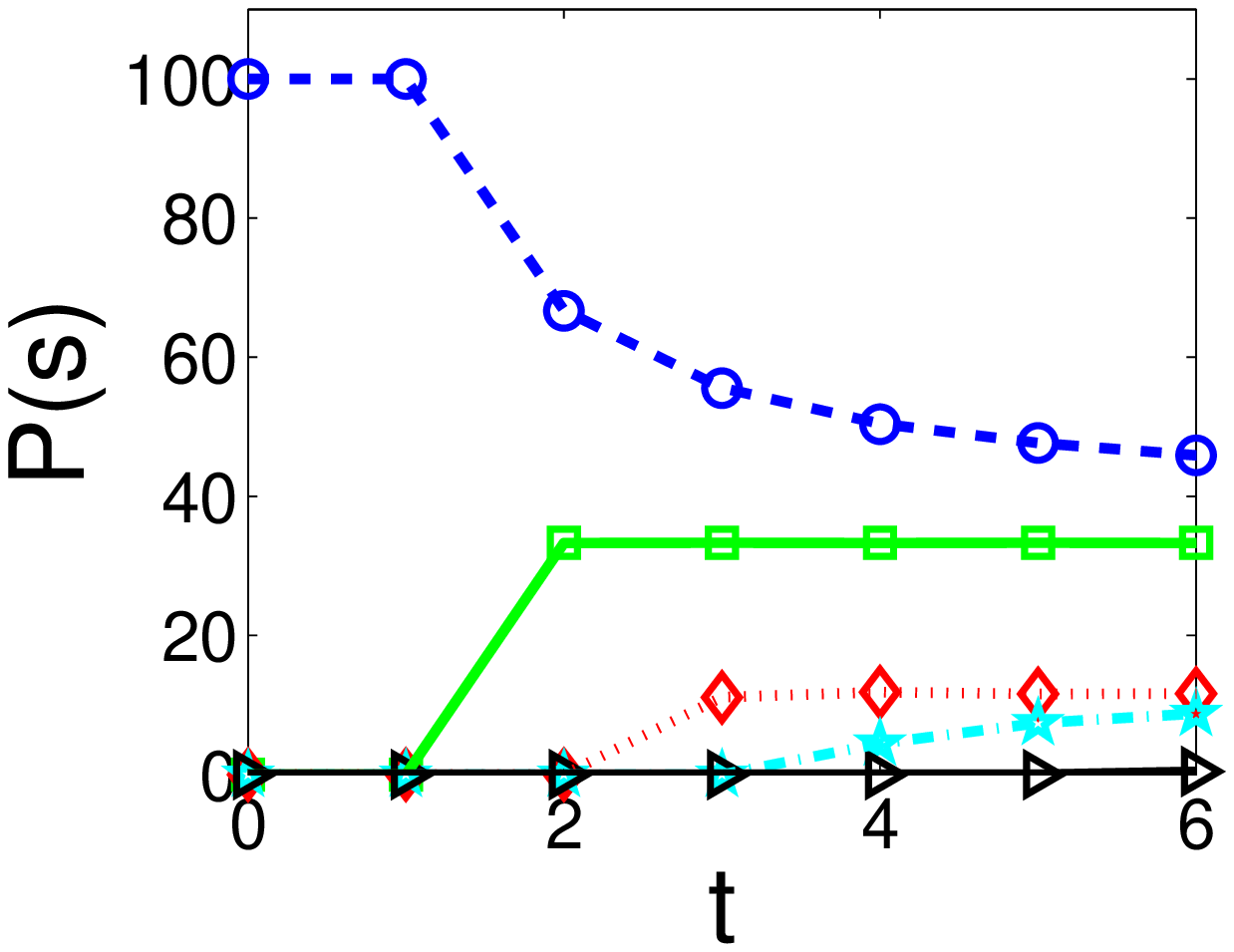}}
\caption{a) Time evolution of the average system size (in percentage), for $C=1$, $L=3$ (blue circles), $L=4$ (green squares), $L=5$ (red rhombus), $L=6$ (blue stars), $L=7$ (black triangles); points are average results from numerical simulations and lines connecting points are theoretical results. b) Time evolution of the average avalanche size (in percentage), for $C=1$, $s=0$ (no-events) (blue circles), $s=1$ (green  squares), $s=2$ (red rhombus), $s=3$ (blue stars), $s=4$ (black triangles); points are average results from numerical simulations and lines connecting points are theoretical results}.
\label{fig8:evol_chain_event_distrib}
\end{figure}

The probability of finding a linear chain of size $L$ at time $t$ can be solved using \textit{Markov chains}.
We define the \textit{transition matrix}, $\textbf{P}$, that contains the probabilities of transition $p_{ij}$ from state $i$ to state $j$ and an initial probability vector $\textbf{u}$ with the probabilities of all the $r$ states at initial time ($\textbf{u}=\left\lbrace s_{1},s_{2},..s_{r}\right\rbrace $). In our particular case  $s_{i}$ refers to a network with $i$ nodes. After $n$ time steps the probability that the system is in state $\textbf{v}=\left\lbrace s'_{1},s'_{2},..s'_{r}\right\rbrace $ can be obtained as the power of the \textit{transition matrix}:
\begin{equation}
\textbf{v}^{(n)} = \textbf{u}\cdot\textbf{P}^{n}
\end{equation}

\noindent where $\textbf{u}= \left\lbrace1,0,0,\cdots \right\rbrace$ for a single-node seed. Note that a linear chain of length $L_{i}$, at time $t$, from a system of length $L_{i-1}$ and from $L_{j}$, for any $j>i$, at time $t-1$ can be obtained. In our work we have studied their first $30$ time units. At stationary state, the average system size (number of nodes) is $5.23$. 
This statistical value is confirmed by numerical simulations (averaged over $10000$ realizations), \textit{i.e.}  $\left\langle L_{C=1}(t=30) \right\rangle = 5.23$. The probability distribution $\textbf{v}$ at any time can be defined for any value of the system size, even for very large number of nodes. For example, the probability of finding a linear chain of size $L=20$ at $t=30$ is about $10^{-9}$. 
Fig.~\ref{fig9a:evol_prop_size_C1} shows the time evolution of the percentage for networks with size $L$, from $L=3$ to $L=7$ (only up to $t=6$ for clarity purpose). Dashed lines correspond to values obtained from the statistics study and dots correspond to average values from numerical simulations. 
From the probability transitions in the Markov chain the average event size distribution can also be estimated. In Fig.~\ref{fig9b:evol_event_distr} the dashed lines represent the theoretical values computed from the statistics study and the dots represent the results from numerical simulations averaged over $10^{8}$ realizations.

This statistical approach to the special case of linear chains can be used to gain a better insight into the event size distribution. Even in this simple case the event size can vary from $1$ to more than a half of the system size at any time, and with this approach we can obtain the probability of an event of any size, and also the probability of having a linear chain of size $L$ at any time. As can be seen in Figure \ref{fig8:evol_chain_event_distrib} the statistical approach and the numerical simulations are in complete agreement.

\section{Conclusions}
\label{Sec:Conclusion}

The characteristic behavior of the SOC dynamics and some of its statistics properties can be analyzed with our simple network model. In this model the system (the network) is maintained out of equilibrium by a constant flux of matter (nodes). The criticality appears due to a stability condition which relates one node's topological property (its degree) with its neighborhood (the average degree). This local condition is associated to a neighborhood's assortativity. This new approach represents one step beyond the Newman's assortative mixing: a node's tendency of linking does not depend on its neighbors' property but on the neighborhood's average property. This assortative mixing by neighborhood's average property should be more suitable for studying social communities networks. An exhaustive study of \textit{neighborhood assortativity} with real and synthetic networks is in progress. We have found that some real networks exhibit positive neigborhood assortativity and null degree assortativity.

In this toy model, the interplay between topology and dynamics drives the system to a self-organized stationary state. The only parameter in the model that controls the system size at the stationary state (without any dynamical variable) is the \textit{buffering capacity} constant $C$.

In order to characterize the SOC dynamics we have performed simulations for different values of the \textit{buffering capacity}. The statistics of events and time intervals between events show distributions with similar exponent ($\gamma = 0.8$) to the ones observed in OFC model \cite{Caruso07}. Moreover, all the distribution plots for different \textit{buffering capacity} constants (and for different system sizes) can be collapsed into an universal curve, indicating that the own dynamics is tuning the phenomena in the same organized way, without external conditions.
The PDF of the ``returns'' (differences between avalanche size at time $t$ and $t+\Delta t$) can be fitted by a \emph{q-Gaussian} curve. The fit exponent $q=2.2 \pm 0.2$ can also be compared with the exponent found in the OFC model \cite{Caruso07}. In general, the model exhibits a SOC behavior with exponents similar to the OFC model.

We have also studied the statistical model for the special case of linear chains ($1\leq C < 4/3$) by means of the \textit{Markov chains}. With this procedure we have obtained the probability of finding the system in a state $s_{i}$ (a network of $i$ nodes) at time $t$ and moreover, it can reproduce the system size distribution obtained from simulations.

By producing small variations which allow cycles and clusters, the model can become a possible representation for some social organizations, such as corporation hierarchy or population organization.

\begin{acknowledgments}

We gratefully acknowledge Miguel \'{A}ngel Ib\'a\~nez, Ram\'{o}n Alonso, Miguel \'{A}ngel Mu\~{n}oz and Juan Carlos Losada for fruitful discussions.
This work was supported
by the project MTM2012-39101 and MTM2015-63914-P from the Ministry of Economy and
Competitiveness of Spain. Large simulations are supported by CESVIMA (supercomputation center of the Technical University of Madrid). 

\end{acknowledgments}


\end{document}